# Tropical Pacific SST influence on seasonal streamflow variability in Ecuador

César QUISHPE-VÁSQUEZ, Sonia Raquel GÁMIZ-FORTIS, Matilde GARCÍA-VALDECASAS-OJEDA,
Yolanda CASTRO-DÍEZ, María Jesús ESTEBAN-PARRA
Departamento de Física Aplicada, Universidad de Granada, Granada, España
cesarqv@correo.ugr.es, srgamiz@ugr.es, mgvaldecasas@ugr.es,
ycastro@ugr.es, esteban@ugr.es

**Abstract**

This study presents a basin-wide assessment about the spatiotemporal variability of streamflows in Ecuador for the period 1979–2015. The influence of the tropical Pacific sea surface temperature (SST) on streamflow variability from February to April (FMA) period, as the months showing maximum streamflow for the wet season in Ecuador, and from June to August (JJA), corresponding to the dry season, was analysed. Firstly, a long-term trend analysis was carried out by applying the Sen's slope estimator and the Mann–Kendall test to monthly streamflow data from 45 gaging stations located in different basins across Ecuador. While the coastal region showed the highest generalized positive trends from July to January, the results for the Pacific Andean area suggested a strengthening of the seasonality, presenting an overall increase in the streamflow for all months except August, September and October, which showed negative trends. Secondly, a singular-value decomposition (SVD) was applied in order to find the main coupled variability patterns between the FMA streamflow and the quasi-coetaneous SST (December to February, DJF), and between the JJA streamflow and the coetaneous SST. The results revealed two main coupled modes for DJF SST/FMA streamflow, the first associated with the canonical El Niño and the second with El Niño Modoki. The latter exerted a major influence on FMA streamflow over most of Ecuador. For JJA streamflow, however, the pattern associated with the traditional El Niño was even more relevant. These results establish the foundations for streamflow modeling in Ecuador based on the Pacific SST, showing the strong response of thestreamflows to different types of El Niño events.

## 1. Introduction

A few decades ago, efforts were initiated in Ecuador to study climate variability and determine the influence of the El Niño/Southern Oscillation (ENSO) on meteorological variables such as precipitation and temperature (Vuille et al., 2000; Pineda et al., 2013; Zubieta et al., 2017; Tobar and Wyseure, 2018; Morán-Tejeda et al., 2016). Ecuador's climate is affected by very diverse factors such as its location near the Intertropical Convergence Zone (ITCZ), the Pacific Ocean, the advection of moisture from the Amazon basin, and the complex topography of the Andes mountain range (Rossel, 1997). The seasonal distribution of rainfall across Ecuador, along with the topography of the Andes, different soil types and land cover, determine the country's availability of water resources (Tobar and Wyseure, 2018). All these factors, in conjunction with the availability of climate data recently collated and published by the competent organisms, mean the study of climate trends and variability across Ecuador are of particular interest (Morán-Tejeda et al., 2016).

Ecuador is usually split into four geographical regions: the Coast, the Andes, the Amazon and the Galapagos Islands (Pourrut et al., 1995; Rossel, 1997). Andean headwater catchments are an important source of freshwater for downstream water users (Urrutia and Vuille, 2009; Roa-García et al., 2011; Molina et al., 2015). The tropical Andean region has been identified as an area of high vulnerability to climate change and hydroclimatic risks (IPCC-SREX, 2012; WFP, 2014). In addition, the rapid growth of certain cities situated in the high Andes, for example Quito, will increase the demand on water resources. However, only a limited number of long-term studies have been conducted in this region regarding the relative importance of the effect of climate change on water resources (Villar et al., 2009; Haylock et al., 2006; Morán-

Tejeda et al., 2016). Moreover, these studies only analyzed rainfall trends and even then they did not agree about their sign, as they were highly dependent on the time period and the gauging stations analyzed (Morán-Tejeda et al., 2016). Additionally, it is important to consider that changes in streamflow may not necessarily be a consequence of precipitation [alone], but also of changes in land use and land cover (Piao et al., 2007). In recent years, land use in Ecuador has changed due to different factors such as population growth, internal migration, land reforms, and increasing agricultural exports, among others (De Koning et al., 1999). To characterize the impacts of land-use on the catchment hydrological response requires, however, a dense hydrometeorological dataset, which is especially complicated in mountain environments (Ochoa-Tocachi et al., 2018). In this context, Molina et al. (2015) showed a different behavior for the trends detected in rainfall and streamflow for a specific location in the Ecuadorian Andean region. Thus, while a long-term increasing trend for rainfall was observed at this location, the streamflow exhibited a decreasing trend. Therefore, streamflow variations in Ecuador could not be necessarily due to long-term changes in precipitation, but may be the result of land cover changes associated with anthropogenic disturbances (Molina et al., 2015). Consequently, there is a need for a detailed basin-wide assessment of streamflow trends that covers the entire country.

Ocean currents have a significant influence on the Ecuadorian coast, particularly the Humboldt Current, which is a key component of the El Niño/Southern Oscillation (ENSO) phenomenon (Rasmusson and Carpenter, 1982). For this area, the studies show that the ENSO has a strong influence on rainfall (Rossel and Cadier, 2009; Tobar and Wyseure, 2018), with an increase in precipitation associated with warm Pacific SST anomalies during El Niño events. During these events, the trade winds weaken along the equator, leading to a weakening of the Walker circulation as atmospheric pressure rises in the western Pacific and falls in the eastern Pacific. This produces an eastward expansion of warm waters from the central Pacific, thereby weakening the Humboldt Current and blocking the equatorial and coastal upwelling (McPhaden et al., 2006; Wang and Fiedler, 2006; Rossel and Cadier, 2009). Atmospheric convection cells and the precipitation associated with the ITCZ shift to the south causes heavy rainfall over coastal Ecuador (Horel and Cornejo-Garrido, 1986; Neill and Jørgensen, 1999; Hastenrath, 2002; Waylen and Poveda, 2002; Poveda et al., 2006).

For more interior areas, the Andes barrier modifies this ENSO influence (Vuille et al., 2000; Villacis et al., 2003), so the Ecuadorian Andes tend to experience the opposite effect to that observed at the coast, with anomalous low precipitation during El Niño events because an anomalous Hadley cell inhibits convection over the high terrain (Vuille et al., 2000; Francou et al., 2004). Several studies have investigated how far the ENSO influence extends inside the country (Bendix et al., 1997; Rossel et al., 1999; Bendix, 2000; Bendix et al., 2003; Morán-Tejeda et al., 2016; Campozano et al., 2016). In general, all the studies showed that positive rainfall anomalies during ENSO mainly affect the region extending from the coastal plain of Ecuador to the western slope of the Andes. Rossel and Cadier (2009) found that the relief of the Andes corresponds to the boundary of the positive ENSO influence. However, there is no clear explanation of how far into the Andes the effects of the ENSO are perceived (Vuille et al., 2000; Morán-Tejeda et al., 2016). Note that for eastern Ecuadorean Andes, evidences of the ENSO impact on the ablation rates of glaciers have also been found (Francou et al., 2004; Veettil et al., 2014).

Therefore, the analysis of the transition of the large-scale influence of ENSO on the Ecuadorian water resources from the coastal plain towards the Andes is necessary (Pineda et al., 2013). To this end, most studies have used different El Niño teleconnection indices. However, there is no agreement among the scientific community about which index best captures the ENSO events (Hanley et al., 2003) or which has the greatest effect over the entire territory (Tobar and Wyseure, 2018; Morán-Tejeda et al., 2016). Most studies either analyzed the precipitation in certain areas of Ecuador (Rossel and Cadier, 2009; Guenni et al., 2017; Pineda et al., 2013) or focused on correlations with specific El Niño indices. The most used teleconnection indices are El Niño 1+2, El Niño 3.4, El Niño 3 and El Niño 4 (Tobar and Wyseure, 2018; Morán-Tejeda et al., 2016; Vicente-Serrano et al., 2017), although the Southern Oscillation Index (SOI) and the Multivariate ENSO Index (MEI) have also been considered (Villar et al., 2009). However, some recent studies on regions neighboring Ecuador have found that the so-called El Niño Modoki (ENM) has a strong influence on precipitation (Ashok et al., 2007; Tedeschi et al., 2013; Córdoba-Machado et al., 2015a, 2015b). We must therefore consider the entire tropical Pacific SST to reach a better understanding of the relationship between the ENSO and water resources in Ecuador.

In 1983 and 1998, Ecuador experienced two significant episodes of torrential rain, high river runoff and flooding associated with the impact of the ENSO; both involved substantial losses of human life and infrastructure. To reduce the human and economic consequences of extreme weather events, the Ecuadorian Government implemented several flood protection projects on some of the country's main rivers (CEDEGE, 1995; Rossel et al., 1999). In Ecuador's neighboring regions the ENSO has a stronger effect on streamflow than precipitation (Poveda et al., 2001). This could be because seasonal variations in streamflow are controlled by large-scale fluctuations in atmospheric circulation patterns (Rimbu et al., 2004, 2005), as hydrological systems integrate, spatially and temporally, basic climate variables such as precipitation (rain and snow), temperature and evaporation from a specific region (Coulibaly and Burn, 2005; Gámiz-Fortis et al., 2010, 2011). So, there are two main reasons driving the growing interest in ENSO-streamflow relationships. Firstly, because the ENSO exhibits a considerable potential predictability that can provide predictions several seasons in advance (Simpson et al., 1993; Gutierrez and Dracup, 2001; Whitaker et al, 2001; Kirtman and Schopf, 1998; Latif et al.,1998, 2001; Goddard and Mason, 2002; Kirtman et al., 2002; Chen and Cane, 2008; Jin et al., 2008), and secondly, for some regions in the vicinity of Ecuador, the tropical Pacific SST represents a suitable tool for predicting seasonal precipitation for the next one to four seasons (Córdoba-Machado et al., 2015b). Therefore, based on the use of Pacific SST as predictor variable, and taking into account its impact on the water resources of specific regions, a relatively high degree of seasonal streamflow predictability could be established (Solow et al., 1998; Córdoba-Machado et al., 2015a, 2015b, 2016). This information could help to mitigate the consequences of both flooding events and droughts. Then, this work is devoted to study the tropical Pacific SST influence on seasonal streamflow variability in Ecuador and attempts to quantify the potential for predicting streamflow from the SST.

Firstly, the long-term trends of Ecuadorian streamflows based on monthly data was carried out. Secondly, the influence of the complete tropical Pacific SST, as a determinant factor in seasonal streamflow variability, was assessed through a coupled singular value decomposition (SVD). Finally, based on the foregoing results, an experiment to reconstruct the streamflow time series was developed for two extreme El Niño events.

## 2. Study Area and Data

Ecuador is situated in the northwest of South America, between Peru and Colombia, covering a latitude of 1.58° N–3.48° S and longitude of 75.28–81.08° W. The area of the present study is shown in Figure 1. Although the Galapagos Islands, which lie approximately 1,000 km to the west of the mainland, form part of Ecuador they are not included in this study. Streamflow data were obtained from Ecuador's National Institute of Meteorology and Hydrology (INAMHI). The data were collected over a period of more than 39 consecutive years from approximately 200 active automatic stations with monthly streamflow time series, but only time series that included at least 85% of the full dataset were considered. Consequently, data for the period 1979–2015 from just 45 stations, distributed across the country (Figure 1), were selected.

Missing data were filled using the R package Amelia II (Honaker et al., 2011). The package performs multiple imputations (Rubin and Schenker, 1986) to fill out incomplete datasets. We considered five imputations to estimate all the missing values for the streamflow time series used in this study (Table 1).

Additionally, streamflow data quality was analyzed to discard any monthly streamflow time series with a nonhomogeneous behavior. In virtue of its nonparametric nature, the Pettitt test (Pettitt, 1979) was used to determine the presence of any inhomogeneities. The Pettitt test provides the year of the breakpoint in the analyzed time series. Subsequently, we used the common area index (CAI, Kundzewicz and Robson, 2004; Hidalgo-Muñoz et al., 2015) to determine whether the abrupt change highlighted by the Pettitt test was due to a change in the natural regime (e.g. a result of dam regulation). This index represents the percentage of common area between the curves of the annual cycles calculated using data from before and after the point of change. Any series presenting a breakpoint with a CAI of less than 50% was considered nonhomogeneous, since the natural streamflow regime was strongly affected after the break (Hidalgo-Muñoz et al., 2015). As a result, the 45 filled streamflow series passed these homogeneity tests and were used in the subsequent analysis.

As shown in Figure 1, the Andes mountain range crossing Ecuador gives the country a complex topography and a range of regional climates. Furthermore, its geographical position on the equator and close to the Pacific Ocean defines the hydrological regimes of Ecuadorian streamflows (Vuille et al., 2000). The annual cycle of precipitation over Ecuador is largely controlled by the meridional migration of the Intertropical Convergence Zone (ITCZ) (Sierra et al., 2015), producing unimodal or bimodal precipitation patterns over specific regions. Traditionally, the continental territory can be categorized into three distinct regions in terms of precipitation (Pourrut et al., 1995; Rossel, 1999): the coast, the highlands or Sierra, and the eastern interior lowlands (the upper Amazon basin). For the westernmost part of the territory, the wet season runs from December to May (accounting for 80% of annual rainfall) and the dry season from June to November.

Recently, Tobar and Wyseure (2017), who used a clustering technique involving monthly rainfall data from 319 stations for the period 1982–2011, also selected a fourth region: the coastal Orographic Sierra, referring to the Pacific side of the Sierra. The classification of seasonal rainfall patterns and their spatial distribution by region, which were key aspects in Tobar and Wyseure (2017), established a relatively uniform type of rainfall for the Amazon area, with the heaviest tropical rains usually occurring in July and August. The coastal region and Pacific side of the Sierra presented the strongest seasonality, with February to April exhibiting the maximum precipitation. In the mountain region (Sierra), the influence of both the coastal and Amazon weather patterns produced two rainy seasons, February to March and June to August. Precipitation in the mountains was generally low intensity but prolonged over time.

In agreement with these results for streamflow in Ecuador, Figure 2 shows the spatial distribution of the seasonal variability of the mean monthly flow. The annual cycles of monthly river flow can be divided into two types: those which peak between February and April, corresponding to the coastal and Pacific Andes regions; and those which peak between June and August, corresponding to the Amazon region. Taking into account the maximum streamflow values observed for the two different flow types, the seasonal averages were calculated from the monthly streamflow time series for the periods from February to April (FMA) and June to August (JJA).

We used the tropical Pacific SST, covering the area 120° E–78.5° W, 30° S–25.5° N, which is taken from the HadISSTv1.1 dataset (Rayner et al., 2003) collected by the Hadley Centre for Climate Prediction and Research (UK Met Office), as the driving variable for streamflow variability. We generated the usual climatological winter, spring, summer and autumn SST fields by averaging the monthly SST anomalies (using the mean and standard deviation for the period 1979–2015) for December-January-February (DJF), March-April-May (MAM), June-July-August (JJA) and September-October-November (SON).

We also employed typical teleconnection indices associated with the ENSO phenomenon published by the US National Oceanic and Atmospheric Administration (NOAA): El Niño 1+2, El

Niño 3, El Niño 4, El Niño 3.4, and the SOI. In addition, other composite indices were used: the Multivariable ENSO Index (MEI, Wolter and Timlin, 1998), which is obtained from six variables related to sea level pressure, winds, sea surface temperature, air temperature and cloudiness; the Trans-Niño Index (TNI, Trenberth and Stepaniak, 2001); and the El Niño Modoki Index (Ashok et al., 2007). The TNI and the El Niño Modoki Indeces are both related to longitudinal SST gradients in the tropical Pacific, and this latter recently considered a key factor impacting on precipitation variability over some areas of South America (Grimm and Tedeschi et al., 2009; Córdoba-Machado et al., 2015a, 2015b, 2016).

Additionally, we also used NCEP/NCAR reanalysis data (Kalnay et al., 1996) to obtain atmospheric fields of velocity potential at 200 hPa and vertical velocity for the levels from 1000 hPa to 100 hPa. This database has a spatial resolution of 2.5 degrees latitude x 2.5 degrees longitude on a global grid and a temporal coverage of 1948 to 2015.

## 3. Methodology

In this study, trends in the monthly streamflow time series were calculated using the nonparametric Sen method (Sen, 1968) to estimate the slope. The statistical significance of the estimated trends was evaluated using the nonparametric Mann–Kendall test (Mann, 1945; Kendall, 1975) at a 0.05 significance level against the null hypothesis of no trend. This test has been widely used in hydrological studies (e.g. Bouza et al., 2008; Hidalgo-Muñoz et al., 2011), and has the advantage of being a rank-based test that is robust to outliers and does not depend on the assumption of a Gaussian distribution of residuals. The modified Mann–Kendall test has previously been used to avoid the potential autocorrelation effect on the significance analysis (Hamed and Rao, 1998).

To assess the total influence of the tropical Pacific SST on streamflow across Ecuador and without limiting it to the regions associated with the teleconnection indices, the singular-value decomposition (SVD) was applied to both seasonal SST and streamflow. This method analyzes the covariability between the tropical Pacific patterns and the streamflow for the different seasons being considered. The SVD technique is a generalization of the diagonalization procedure performed in the principal component analysis (Preisendorfer, 1988) of nonsquare or nonsymmetrical matrices. This method was first used in a meteorological context by Prohaska (1976) and later by Lanzante (1984). As applied by Bretherton et al. (1992), the SVD is performed on the cross-covariance matrix of two fields. It isolates the linear combinations of variables within the linearly related fields, hence maximizing the covariance between them. The covariance matrix is constructed from the anomalies of the variables with respect to their mean values. The eigenvectors resulting from the diagonalization of this rectangular matrix are called "right" and "left" singular vectors (similar to empirical orthogonal functions, EOFs) and correspond to each of the variables respectively. They are ranked according to their corresponding eigenvalue so that the first pair of eigenvectors gives the maximum amount of total square covariance (SCF) between the two fields (Björnsson and Venegas, 1997). The SCF is an efficient means of comparing the relative importance of the modes in the decomposition (Bretherton et al., 1992). By projecting the original field on the singular vector, we can calculate the expansion coefficient time series, which represent the temporal variability of the associated spatial patterns (similar to the principal component time series in EOF analyzes). The correlation between the expansion coefficient time series for each variable, known as the relative strength of coupling (SF), measures the degree of the correlation between the coupled patterns. The heterogeneous correlation maps represent the correlation coefficients between the values of each grid point of a field and the expansion coefficients of the other field mode. In this study, this indicates the capacity to predict the seasonal streamflow time series based on knowledge of the seasonal SST expansion coefficient. The significant variability modes revealed by the SVD were selected using the method proposed by North et al. (1982). A more detailed explanation of the technique can be found in Björnsson and Venegas (1997).

In the present work, the SVD technique was applied to quasi-coetaneous (DJF SST and FMA streamflow) and coetaneous (JJA SST and JJA streamflow) values for wet and dry seasons, respectively. We used quasi-coetaneous values for the wet season because ENSO events are often close to or just past the peak in DJF, but may have significantly dissipated by FMA (Recalde-Coronel et al., 2014).

We performed a composite analysis of the velocity potential at 200 hPa and the vertical velocity profile averaged between a latitudinal band of 5° N–5° S to explore the physical mechanism driving the influence of the tropical Pacific DJF SST on FMA streamflow in Ecuador. Therefore, we considered the extreme events associated with the different types of El Niño that were found to be significant modes of coupled variability in relation to the streamflow. This composite analysis was carried out using only those years when the different types of El Niño did not coexist.

Finally, having established the link between the Pacific SST and streamflow, a regression analysis was carried out in order to reconstruct the seasonal streamflows from the SST. We also analyzed the consistency between the patterns of reconstructed and observed streamflow anomalies through the spatial correlation coefficients and the percentage of variance explained (Córdoba-Machado et al., 2015a).

## 4. Results

### 4.1 Streamflow trends

The results of the monthly streamflow trend analysis are shown in Figure 3 and Table 2. Twelve monthly trend slopes were estimated for each station. The percentage of positive and negative trend slopes and the significant ones by month are presented in Table 2. Monthly trends provided overall positive slopes with primarily positive coefficients, but for some stations no positive trends were observed at the 95% confidence level, especially from March to June. Spatially, there was a general increase in streamflow for all months except August, September and October in the Andean region, when the percentage of significant negative trends included 13% in August, 29% in September and 20% in October (Table 2). For this region, however, November to February showed increasing trends that were more significant for January and February. The coastal region presented the highest generalized positive slopes from July to January, while there were few significant trends from February to June. Only two stations, one on the north coast and the other in the middle of the coastal region, showed significant streamflow attenuation for September and October. For the Amazon region, only one station displayed a significant decreasing trend in September.

To determine whether these trends were supported by rainfall trends, an analysis of monthly precipitation trends was carried out (Figure S1 in the supplementary material). Increasing and decreasing precipitation trends were generally in agreement with streamflow trends, although several discrepancies are apparent depending of the location. Overall, the observed trends in streamflow are principally the result of long-term changes in precipitation.

**4.2 Quasi-coetaneous SST-streamflow coupled variability modes in the wet season**

An SVD analysis was performed on the complete 1979–2015 time series of tropical Pacific DJF SST and FMA streamflow data for Ecuador. This revealed that the first significant SST mode corresponded to the traditional El Niño phenomenon (Figure 4, upper left column), usually characterized by high positive SST anomalies over the eastern Pacific Ocean and weak negative anomalies over the western Pacific Ocean. Table 3 and Figure 5 show the relationship between the expansion coefficients of the DJF SST modes and the teleconnection indices associated with the El Niño phenomenon. The most significant correlations with the expansion coefficients for the first DJF SST mode appeared with the more common El Niño indices: 0.97 with El Niño 3 and El Niño 3.4, -0.87 with the SOI and 0.83 with El Niño 1+2 and El Niño 4 (Table 3). From a running correlation analysis (Figure 5), we can conclude that the correlations remained stable throughout the period. The first SST mode accounted for 35.4% of the squared covariance fraction between the DJF SST and FMA streamflow, with a correlation between the fields of 0.77. The heterogeneous correlation map associated with this SST mode (Figure 4, left center), showed an important number of locations with significant positive correlations in the coastal region and negative correlations in the Andes.

The second coupled mode (Figure 4, top right), corresponding to 25.2% of the squared covariance fraction between the SST and the streamflow, showed a significant coupling, presenting a correlation between the expansion coefficients of both fields of 0.77. The pattern depicted for this second SST mode is spatially similar to the mode associated with the ENM (Ashok et al., 2007). From Figure 4, the second mode was shaped like a boomerang and presented a core of positive SST values in the central Pacific Ocean surrounded by negative SST anomalies in the eastern and western Pacific Ocean, which was very similar to the ENM pattern. Table 3 shows this SST mode presented the highest correlation values with the indices associated with ENM activity (-0.92 with the TNI and 0.85 with the ENM index), which also passed the stability running correlation test (Figure 5). However, the more common El Niño indices presented unstable correlations with this second mode. The heterogeneous correlation map for streamflow in Ecuador showed generalized negative correlations over the coast and Andes, indicating a clear decrease of streamflow during ENM events.

We performed a composite analysis to obtain atmospheric teleconnection patterns that could help explain the negative and positive correlations observed between the Pacific DJF SST modes and the FMA streamflow in Ecuador. Extreme ENSO events were selected based on El Niño 3 and El Niño Modoki indices from years in which the winter had an index value equal to or greater than a standard deviation of 0.5 (Table 4). The correlation coefficient between these two teleconnection indices (Niño 3 and Niño Modoki) during the period 1979–2015 is 0.36, which shows some degree of relation between the two phenomena. However, different authors (Ashok et al., 2007; Kug et al., 2009; Shuanglin and Qin, 2012) considered El Niño Modoki to be a different phenomenon to the conventional ENSO event, while others have proposed that El Niño Modoki is a nonlineal evolution of the ENSO (Takahashi et al., 2011). We analyzed the DJF anomaly maps for the velocity potential fields at 200 hPa, the vertical velocity averaged between the 10° N–10° S latitudinal band, and the tropical Pacific SST averaged for the selected years (Figure 6). The velocity potential at 200 hPa shows information regarding the intensity of atmospheric circulation, which reflects the convergence/divergence processes (Weng et al., 2007).

Figure 6a (left) shows the SST pattern associated with the canonical El Niño events. The anomalous warming extending from the South American coast to the central Pacific is associated with intensification of the divergent flux around 120°W in the equatorial Pacific at high levels (Figure 6b, left). This is accompanied by ascendant vertical movements in the troposphere between 180° E and 80° W and descendent movement over northern South America (Figure 6c, left). There is also descendent movement over Indonesia and Australia during the canonical El Niño events related to the convergence pattern observed at 200 hPa; this is associated with the well-known displacement of the Walker circulation during ENSO years (Weng et al., 2007, 2009). This situation helps explain the significant increase in precipitation, and therefore in the streamflow, in the west of Ecuador during El Niño events.

During the extreme events selected for El Niño Modoki, positive SST anomalies were noted in the central Pacific, with more moderate values than for the canonical El Niño, along with negative SST anomalies in the west and east of the tropical Pacific (Figure 6a, right). This was associated with a decrease in the divergent flow at 200 hPa and descendent movement over Ecuador (Figures 6b and 6c, right). This situation contributes to explain the significant decrease in precipitation in Ecuador, and therefore in the streamflow, associated with El Niño Modoki events.

**4.3 Coetaneous SST-streamflow coupled variability modes in the dry season**

We conducted an SVD analysis for the Ecuadorian dry season on the complete 1979–2015 time series for the tropical Pacific JJA SST and JJA streamflow. During this season (Figure 7), the first and second coupled modes accounted for 39.6% and 15.7% of the squared covariance fraction respectively, between the fields of tropical Pacific SST and streamflow, giving a cumulative total of 55.2% of the squared covariance fraction. The first mode of SST in JJA (Figure 7, top left) showed a similar structure to the canonical El Niño phenomenon, while the second mode (Figure 7, top right) featured a similar pattern but shifted to the west. The correlation analysis (Table 5 and Figure 8) between these two SST modes and the teleconnection indices showed that the first JJA SST mode presented the highest correlation with the El Niño 1+2 index (a stable and significant correlation value of 0.97 throughout the entire period, Figure 8a), while the second JJA SST mode presented a maximum correlation with the El Niño 4 index (a stable, significant value of 0.94, Figure 8b). The heterogeneous correlation map associated with the first mode (Figure 7, left center) showed generalized positive correlation values for the coastal and Andes regions. The heterogeneous correlation map for the second mode revealed some negative correlations along the Andes. It is noteworthy that fewer stations showed a significant correlation than that observed in FMA, which means the tropical Pacific SST has a weaker impact on streamflow during the dry season.

**4.4 Reconstruction of seasonal streamflow in Ecuador**

Multivariate regression analysis was used to evaluate the feasibility of the different types of El Niño to reconstruct the FMA and JJA streamflow series in Ecuador. The independent variables taken for this calculation were the expansion coefficient time series obtained from the first two SST modes that were derived by the SVD analysis of the tropical Pacific SST and streamflow anomalies. The use of these two series instead of the El Niño indices has the advantage of preventing the high collinearity between independent variables in the regression models.

For FMA streamflow, Figure 9a shows the sites with significant correlation between the original and reconstructed streamflow series. While a few stations presented significant correlations when the data was reconstructed from the first DJF SST mode, this number increased sharply when the second mode was used, reaching values of around 0.7–0.8 in some cases. When the reconstruction involved both DJF SST modes the correlation values increased to around 0.8–0.9 for some sites on the coast and in the Pacific Andes. In line with this result, the root mean square error (Figure 9b) exhibited lower values when the two modes were included in the regression models, with values below 10% at many sites in the Pacific Andes. However, some stations on the coast featured errors of around 100%.

A reconstruction experiment was carried out at different sites along the coast and Pacific Andes to highlight the importance of the second tropical Pacific DJF SST mode of variability (associated with the ENM phenomenon) in terms of FMA streamflow. Figure 10 shows the observed streamflow time series, the reconstructed series employing the expansion coefficients for each SST mode separately, and the reconstructed series using a combination of the first two SST modes at each of the six selected sites. The strong contribution of the second DJF SST mode is evident for all six localities, leading to the conclusion that this mode, which is associated with the ENM, is a key factor in the FMA streamflow reconstruction process. While the first DJF SST mode, associated with the traditional El Niño, presented a very limited or no contribution (as in the case of station H0338), the second DJF SST mode captured a large proportion of FMA streamflow variability. In any case, the joint contribution of both modes improved the result of the reconstruction, as collectively they produced the highest correlation between the observed and reconstructed FMA streamflows.

Figure 11a shows the sites with a significant correlation between the observed and reconstructed JJA streamflow series. The first JJA SST mode, which is associated with the El Niño 1+2 index, was dominant for this seasonal streamflow and reached correlation values of around 0.7–0.8 for stations located in the coastal region. However, only a small number of stations in the Pacific Andes presented a significant correlation when the reconstruction used data from the second JJA SST mode. The reconstruction incorporating contributions from both SST modes increased the correlation values to around 0.8–0.9 for some sites in the coastal and Pacific Andes regions. The root mean square error (Figure 11b) presented low values (below 10%) for most of the stations, although the reduction of the error using the two expansion coefficient series of SST seems to be more modest than for the FMA streamflow.

Figure 12 shows the observed JJA streamflow series and the reconstructed series using the first two modes of JJA SST for six sites. In all cases, the first JJA SST mode, which is highly correlated to the El Niño 1+2 index, was the dominant factor influencing the streamflow. Although the second JJA SST mode appeared to have a very low skill to reproduce the streamflow alone, when it was included in the regression model the reconstructed series was able to better capture the streamflow variability, particularly for some of the highest streamflow values. This resulted in a greater correlation between the modeled and observed streamflow series.

Furthermore, we studied the coherence of the reconstructed streamflow values for 1997 and 1998, which were considered years with strong summer (JJA) and winter (DJF) canonical El Niño events, respectively, that exerted a strong impact on Ecuador's climate. To this end, we analyzed the streamflow anomaly patterns by comparing the observed FMA and JJA streamflow anomalies during 1998 and 1997, respectively, against the reconstructed anomalies using the coupled SST modes.

Figure 13 presents the observed (Figure 13a) and reconstructed (Figure 13b) JJA streamflow anomaly maps for 1997. Note the high degree of similarity between the original and reconstructed maps using both SST modes. This result indicates the contribution of the first and second modes of tropical Pacific JJA SST on the JJA streamflow in 1997. Similarly, Figure 13 shows the observed (Figure 13c) and reconstructed (Figure 13d) FMA streamflow anomaly maps for 1998. Again, the contribution of the first two tropical Pacific DJF SST modes when reconstructing the FMA streamflows is shown by the strong spatial consistency among the observed and simulated maps.

## 5. Discussion and concluding remarks

In terms of climate trends, from the streamflow data we can distinguish two well-differentiated regions in Ecuador: the coastal and Pacific Andean regions. Just a few stations are available in the Andean Amazon region and nonsignificant trends were observed in most months. The coastal area showed the highest generalized positive trends, reaching values of above 2% from July to January, while few significant trends were noted from February to June. These positive streamflow trends agree with the ITCZ narrowing detected in recent decades using satellite observations and reanalysis data, which results in increasing precipitation trends that are especially pronounced in the core of the Pacific ITCZ (Byrne et al., 2018). There was a prevalence of positive trends in the Pacific Andean region with an overall increase in streamflow for all months except August, September and October, when significant negative trends, greater than -2% in September, were reached in some locations. This fact is pointing a strengthening of the seasonality in this region; that is, several dry season months are getting drier and rainy season months, from December to February, are getting wetter, which subsequently places greater strain on water resources during the dry period. These findings agree with the results obtained by Tobar and Wyseure (2018) from an analysis of monthly precipitation trends in Ecuador. They classified the country into four areas in terms of rainfall: the coast, the Sierra, the Amazon and the coastal Orographic Sierra. They found trends for the Sierra cluster wherein rainfall declined in September and October, while December, February, March, April and June all showed increases. Morán-Tejeda et al. (2016), working on annual scale and precipitation during the wet (December to May) and dry (June to November) seasons, reported increased annual rainfall trends in the Andean region due to increasing trends in the wet season, which also agree with our results. However, some discrepancies are apparent with respect to our results for the coastal region, as Morán-Tejeda et al. (2016) did not observe any specific trends at these sites.

The comparison of trend analyses for streamflow and precipitation in Ecuador shows that observed changes in streamflow for 1979–2015 are mainly the result of long-term changes in rainfall, although several discrepancies can be noted depending on the location, which is probably a result of land-cover changes. This result agrees with that obtained by Molina et al. (2015), who suggested that long-term changes in streamflow for the Andean region are probably due to anthropogenic disturbances such as land-cover changes. However, note that their study only involved a streamflow time series in the Pangor catchment and the detected trend was nonlinear.

An SVD analysis was used to determine the predictive skill of the tropical Pacific SST for seasonal streamflows in Ecuador. The relation with the quasi-coetaneous DJF SST was studied for FMA streamflow, while the coetaneous JJA SST was used for JJA streamflow.

For FMA streamflow, the first DJF SST mode found was associated with the canonical El Niño phenomenon. For this SST mode, the highest, temporally stable correlation values corresponded to the El Niño 3 and El Niño 3.4 teleconnection indices. The first DJF SST mode presented significant positive correlations with the FMA streamflow in the coastal region, while some negative correlations were apparent in the Andes region. In line with Rossel and Cadier (2009), this quasi-coetaneous relationship observed in the coastal region reflects the significant influence of the ENSO at the end of the rainy season. Our result also coincides with the findings reported by Recalde-Coronel et al. (2014), who analyzed the predictability of FMA rainfall in coastal and Andean Ecuador. They found a similar spatial pattern of coupled SST-streamflow that described the effects of the canonical ENSO in the eastern tropical Pacific on Ecuador's coastal rainfall, while some stations in the northeastern Andes showed a contrasting rainfall response. Similarly, Morán-Tejeda et al. (2016) also found significant positive correlations between the El Niño 1+2 index and rainfall stations located near the coast, while they reported negative correlations between El Niño 3.4 and stations in the Andes.

The second DJF SST mode, calculated by SVD, presented a spatial pattern associated with the ENM. Only El Niño teleconnection indices, which take into account the temperature gradient between the eastern and western tropical Pacific, exhibited significant and stable correlations with this second DJF SST mode. Although both SST modes had a similar coupling strength (around 0.77) with their corresponding FMA streamflow modes, the heterogeneous correlation maps revealed that the second SST mode had a wider relationship with the streamflow variability than the canonical El Niño and presented significant negative correlations with the FMA streamflow anomalies over the coastal and Pacific Andes regions. This indicates that a significant decrease in FMA streamflow occurs across the country during ENM events.

It is worth noting, however, that canonical DJF El Niño events have a much limited influence on FMA streamflow, with only some sites in the coastal area showing significant correlations, reflecting an opposite response in streamflow with respect to the two types of ENSO events. Some authors have analyzed the influence of the canonical El Niño over precipitation in Ecuador (Morán-Tejeda et al., 2016; Recalde-Coronel et al., 2014; Villar et al., 2009; Rossel, 1997; Rossel et al., 1999; Rossel and Cadier, 2009), identifying characteristics that generally concur with the results of this study. However, our study is the only work to analyze the impact of the ENM on Ecuadorian water resources. The results from the composite analyses of the velocity potential fields at 200 hPa and the vertical velocity averaged between the 10° N–10° S latitudinal band for the extreme canonical El Niño and El Niño Modoki events agree with the atmospheric patterns linked to the two types of El Niño (Ashok et al., 2007; Tedeschi et al., 2013). So we can conclude that differences in Walker circulation during the two types of ENSO are responsible for a significant portion of rainfall differences in Ecuador, and therefore of streamflow. Additionally, other important factors that could explain differences in precipitation and therefore streamflow are the orography (Poveda et al., 2011; Córdoba-Machado et al., 2015a), the latitude of the ITCZ, the zonal and meridional components of the upper atmosphere wind and air humidity (Rossel and Cadier, 2009). The anomalously heavy rainfall in the coastal area of Ecuador associated with the canonical El Niño is known to be the result of strong positive SST along Ecuador's coast and an equatorward expansion and intensification of the ITCZ over the eastern Pacific (Vuille et al., 2000). During El Niño phases, unusually warm eastern equatorial Pacific waters promote convection in the overlying atmosphere, while the ITCZ and associated precipitation anomalies shift further south in comparison with the regular seasonal cycle (Bendix and Bendix, 2006).

For the dry season, the SVD analysis also revealed two coupled variability modes between the JJA SST and JJA streamflow. The first JJA SST mode is strongly correlated with the El Niño 1+2 index (stable over time), while the second one presents the greatest stable correlation value with the El Niño 4 index ($r$ = 0.94). However, these two SST modes exhibited an opposite coupling with streamflow. Positive JJA SST anomalies in the El Niño 1+2 region are associated with overall increases in streamflow in the coastal region, while streamflow decreases are observed in the Pacific Andes when these SST anomalies shift westwards over the El Niño 4 region. These results are in accordance with those obtained by Villacis et al. (2003), Francou et al. (2004) and Morán-Tejeda et al. (2016). These last authors found little correlation between precipitation in the Ecuadorian Andes and the El Niño 1+2 index, while the signal of El Niño 3.4 resulted more evident. They concluded that precipitation variability at the rainfall stations placed in the Andes could be partially explained by El Niño 3.4, especially during July and August.

Our reconstruction of seasonal streamflow time series for different locations revealed the importance of the tropical Pacific SST when forecasting streamflow across Ecuador. However, our study did not contemplate other variability modes and physical mechanisms that may be associated with local effects, but which could significantly influence seasonal streamflow in Ecuador (Recalde-Coronel et al., 2014; Morán-Tejeda et al., 2016).

The results obtained from the present work highlight the importance of studying the different types of El Niño events in order to model seasonal streamflows across Ecuador. The ability to reconstruct FMA streamflow in Ecuador is dominated by the El Niño Modoki phenomenon, which is very important for both the coastal and Pacific Andean regions. For JJA streamflow, a more traditional El Niño is the dominant factor to the west of the Andes. In summary, our results constitute the first step towards modeling seasonal streamflow in Ecuador using the main modes of variability of tropical Pacific SST as predictors instead of the teleconnection indices associated with the ENSO phenomenon. Note that the teleconnection indices are representations of the ENSO phenomenon that may not necessarily include SST anomalies in certain areas of the tropical Pacific, which could be essential in determining streamflow anomalies across Ecuador.

**Acknowledgments**

This study was financed by the Spanish Ministry of Economy, Industry and Competitiveness, with additional support from the European Community Funds (FEDER), projects CGL2013-48539-R and CGL2017-89836-R. Streamflow datasets were kindly provided by Ecuador's National Institute of Meteorology and Hydrology (INAMHI). We thank two anonymous referees and associate editor whose comments improved the paper.

**Figure captions**

Figure 1. Location of Ecuador showing the area of study, elevation and spatial distribution of hydrological stations analysed.

Figure 2. Spatial distribution of gauging stations with typical annual streamflow cycles across Ecuador. Units are $m^3s^{-1}$. Light green corresponds to the stations with maximum streamflow in FMA and dark green those with maximum streamflow in JJA. On the right, the streamflow for two stations taken as examples (marked with red squares on the map).

Figure 3. Monthly streamflow trends for the period 1979–2015, given in percentages with respect to the average streamflow. The significance of the trend calculated by the Mann–Kendall test is marked by solid triangles (for a significance level of 0.05).

Figure 4. Main DJF SST modes (top), heterogeneous correlation maps (middle), and standardized expansion coefficient series (bottom) for DJF SST (red) and FMA streamflow (blue), determined from an SVD analysis between quasi-coetaneous fields of the tropical Pacific DJF SST and FMA streamflow in Ecuador from 1979–2015. The heterogeneous maps only include significant values with a 95% confidence level.

Figure 5. Running correlations (15-year windows) between the standardized expansion coefficient series of: a) the first DJF SST mode, and b) the second DJF SST mode, and the different DJF teleconnection indices associated with the El Niño phenomenon. Dashed lines represent the threshold of the 95% confidence level.

Figure 6. DJF anomaly maps obtained from the composite analysis for extreme canonical El Niño and El Niño Modoki events, for the fields of (a) tropical Pacific SST, (b) velocity potential at 200 hPa (x $10^6$ $m^2s^{-1}$), and (c) vertical velocity (x $10^{-3}$ Pa $s^{-1}$) averaged over 5° N–5° S.

Figure 7. As for Figure 4, but for tropical Pacific JJA SST and JJA streamflow.

Figure 8. As for Figure 5 for a) first, and b) second JJA SST modes, and the different JJA teleconnection indices.

Figure 9. (a) Significant (95% confidence level) correlation coefficients between the original and reconstructed series of FMA streamflow obtained through multivariate regression analysis using the expansion coefficients associated with the first two modes of the tropical Pacific DJF SST. (b) Root mean square error (%) of the reconstructed streamflow series with respect to the observed series for the FMA period at each site.

Figure 10. FMA streamflow series ($m^3$ $s^{-1}$) at six different sites (red dots on map of Ecuador), featuring the observed (black line) and reconstructed streamflow series using DJF SST mode 1 (green line), mode 2 (blue line), and both (red line). The values inside the squares indicate the correlations between the observed and reconstructed series using one (mode 1 in green, mode 2 in blue) or two (modes 1+2 in red) expansion coefficient series. Significant correlation values at the 95% confidence level are indicated in bold.

Figure 11. As for Figure 9, but for JJA streamflow.

Figure 12. As for Figure 10, but for JJA streamflow.

Figure 13. Observed (left) and reconstructed (right) streamflow anomaly (%) maps for JJA 1997 (top row) and FMA 1998 (bottom row).

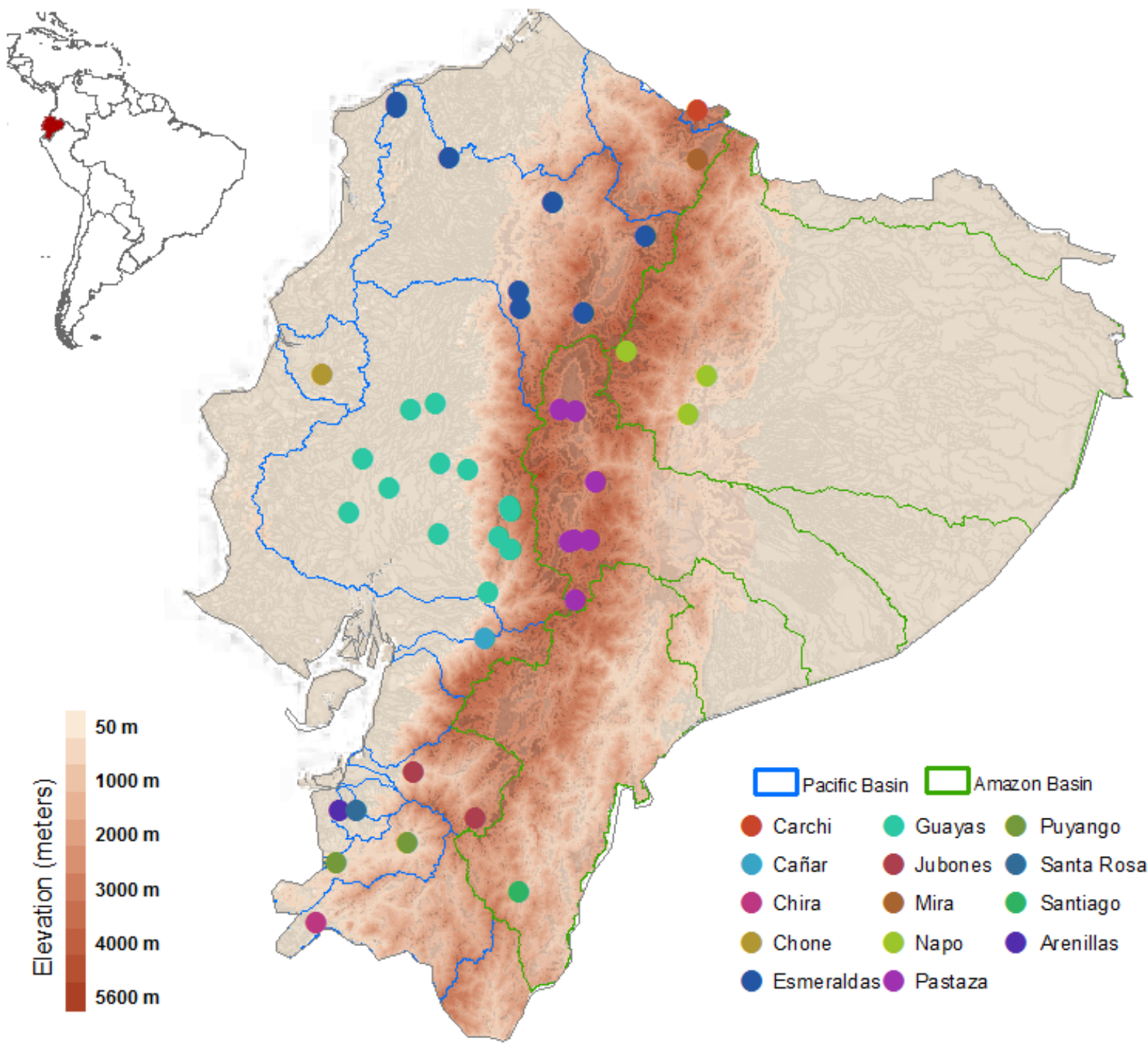

JOC_6047_figure1.png

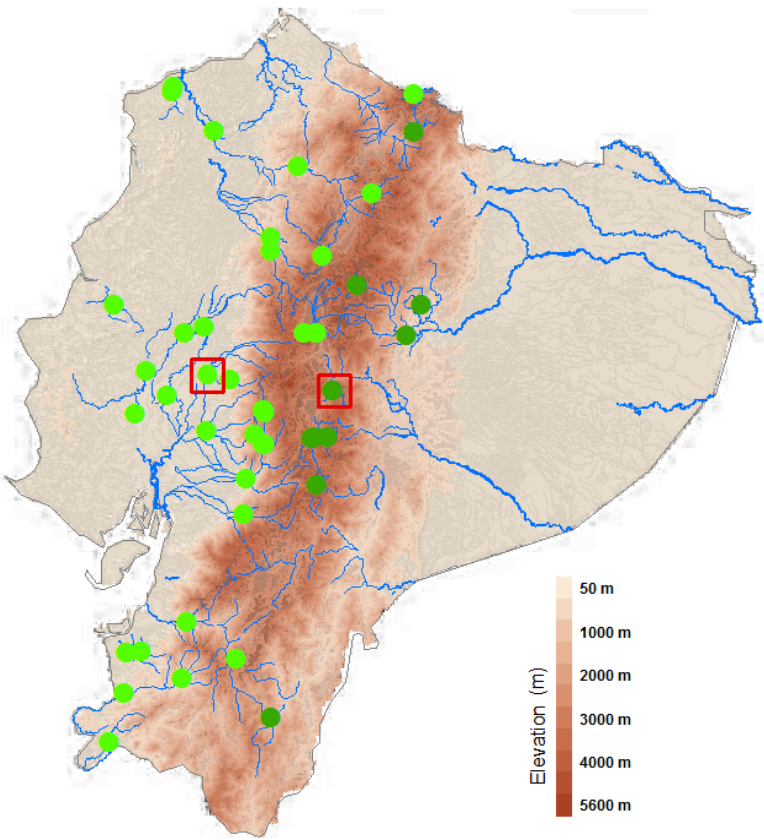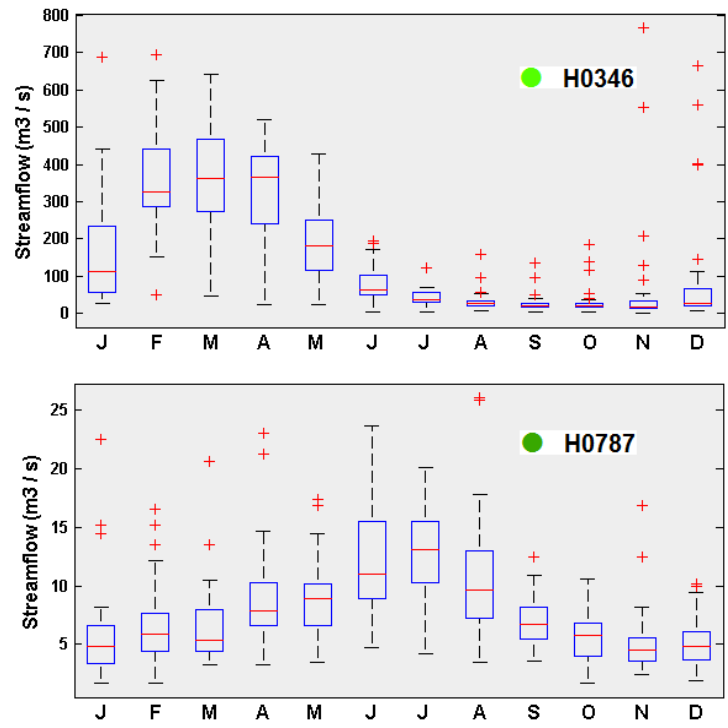

JOC_6047_figure2.png

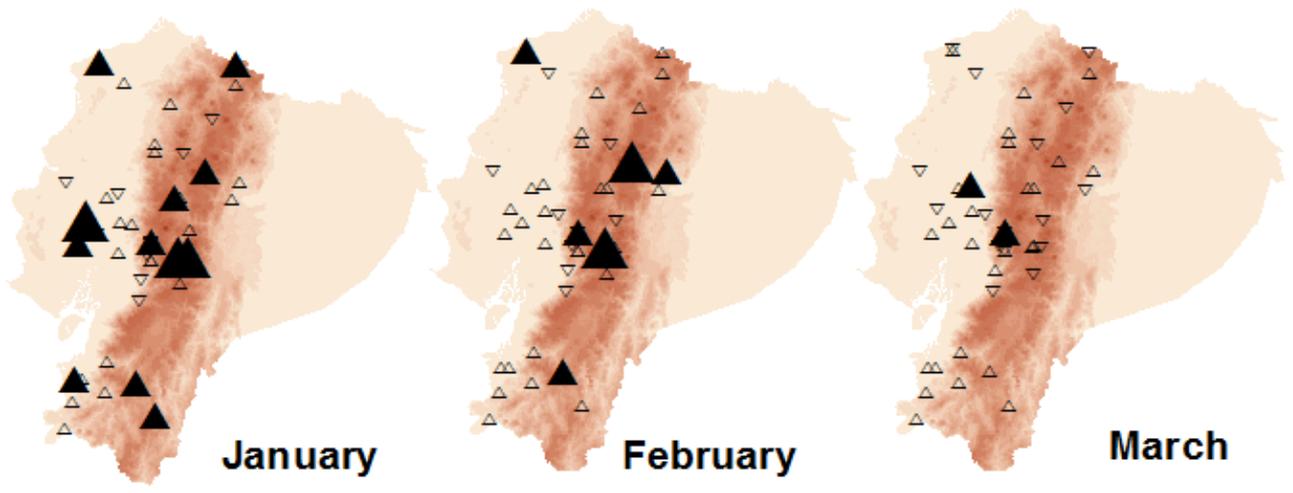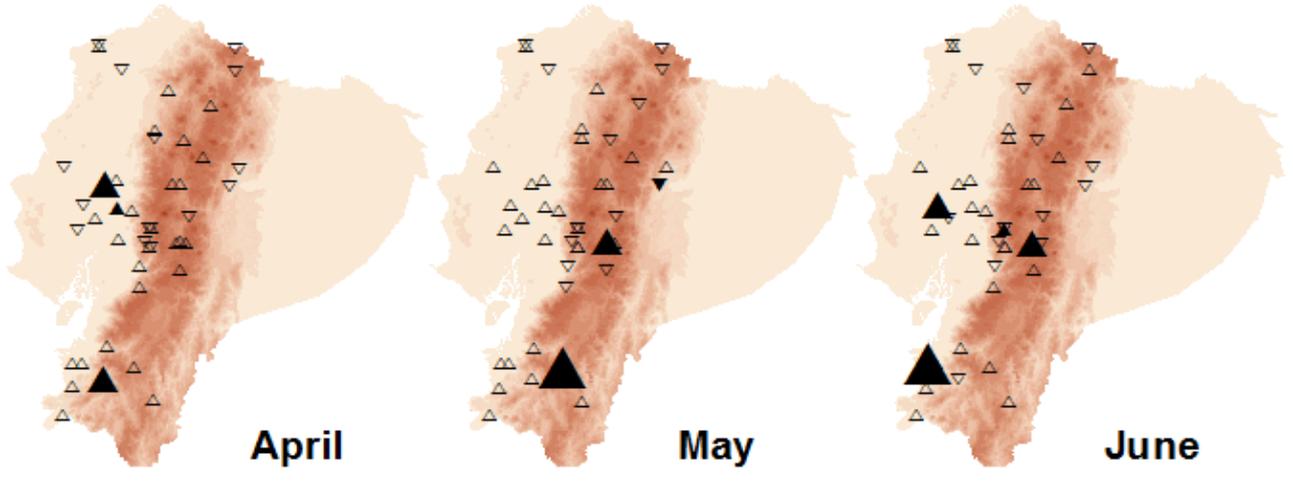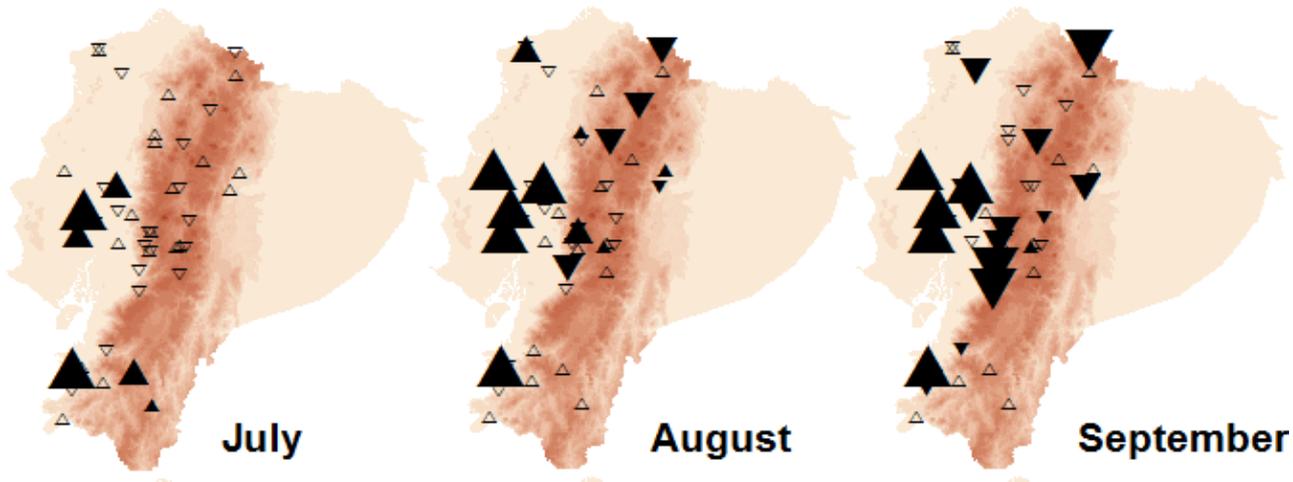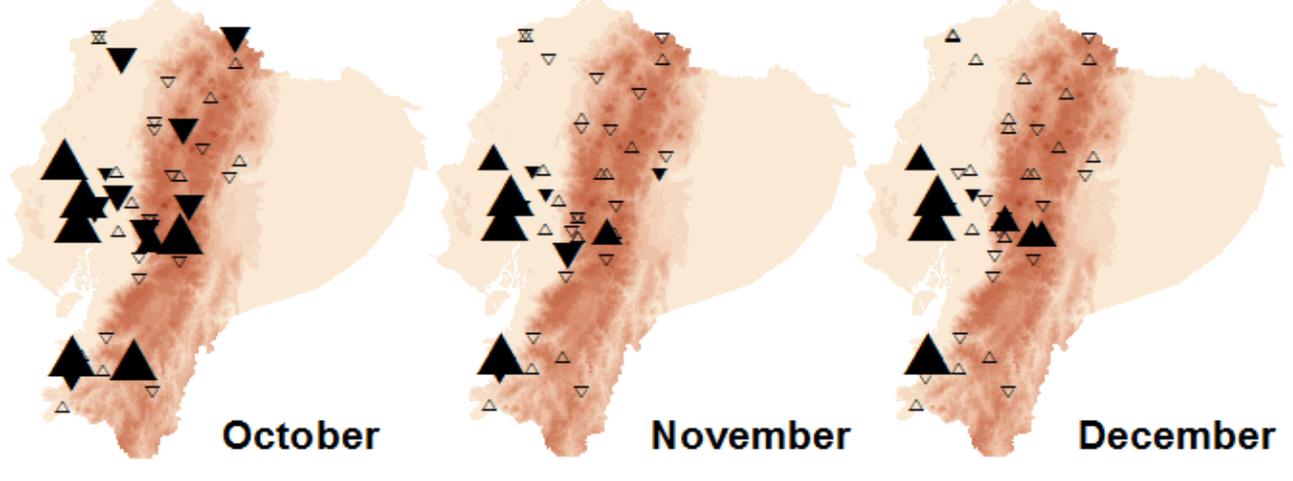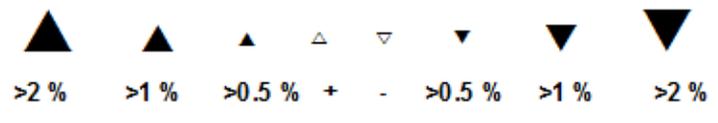

JOC_6047_figure3.png

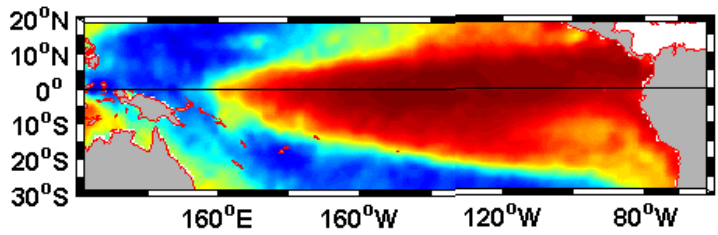
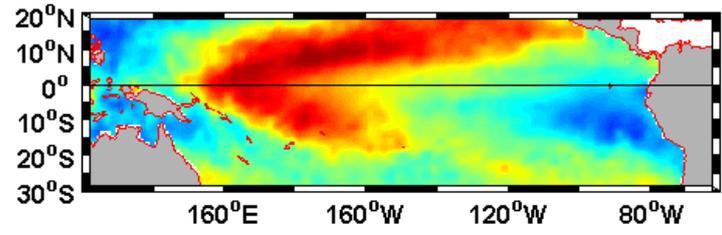
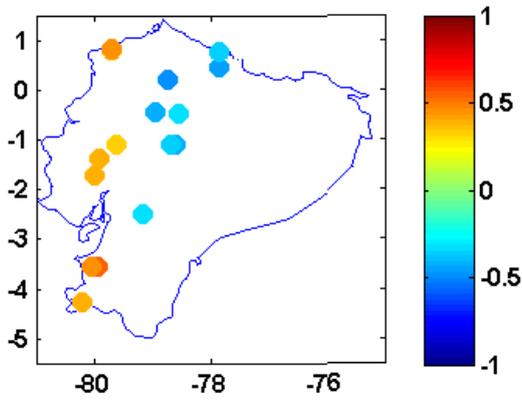
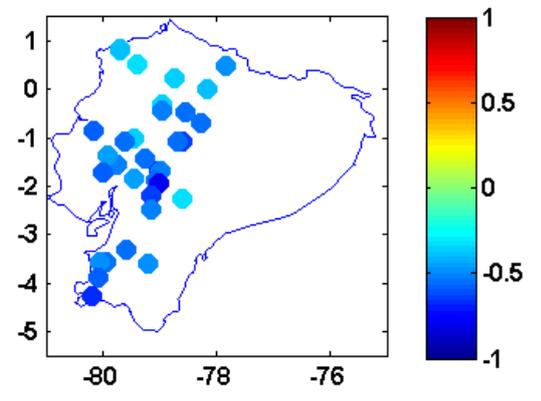
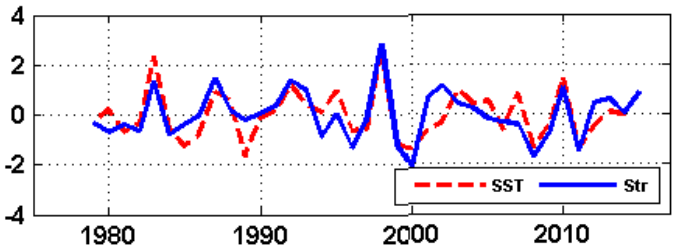
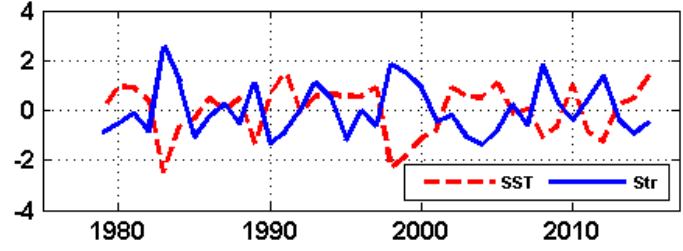

JOC_6047_figure4.png

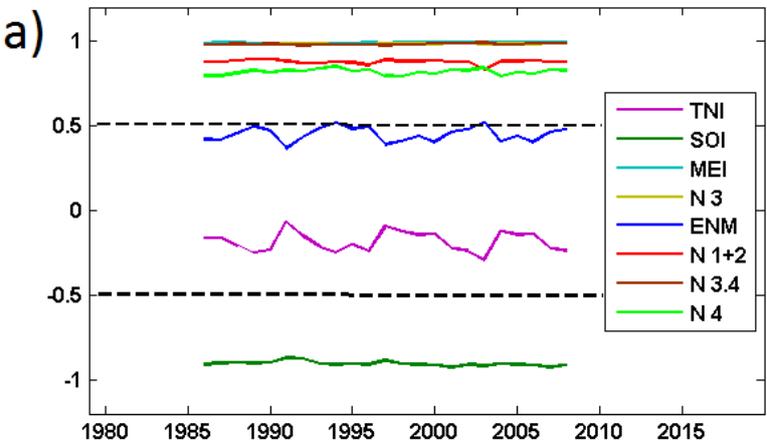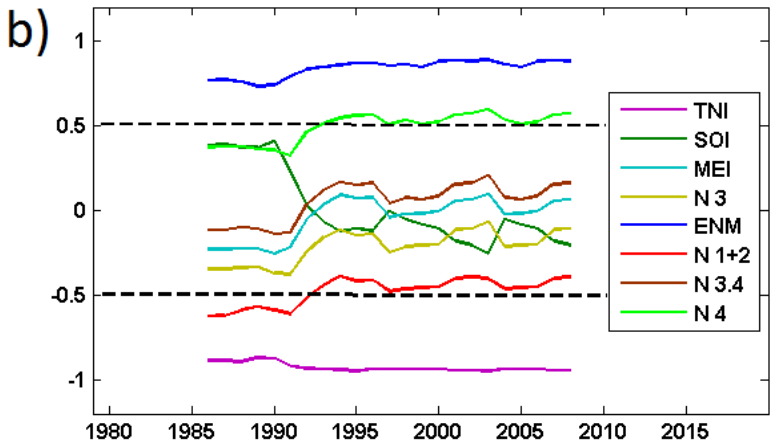

JOC_6047_figure5.png

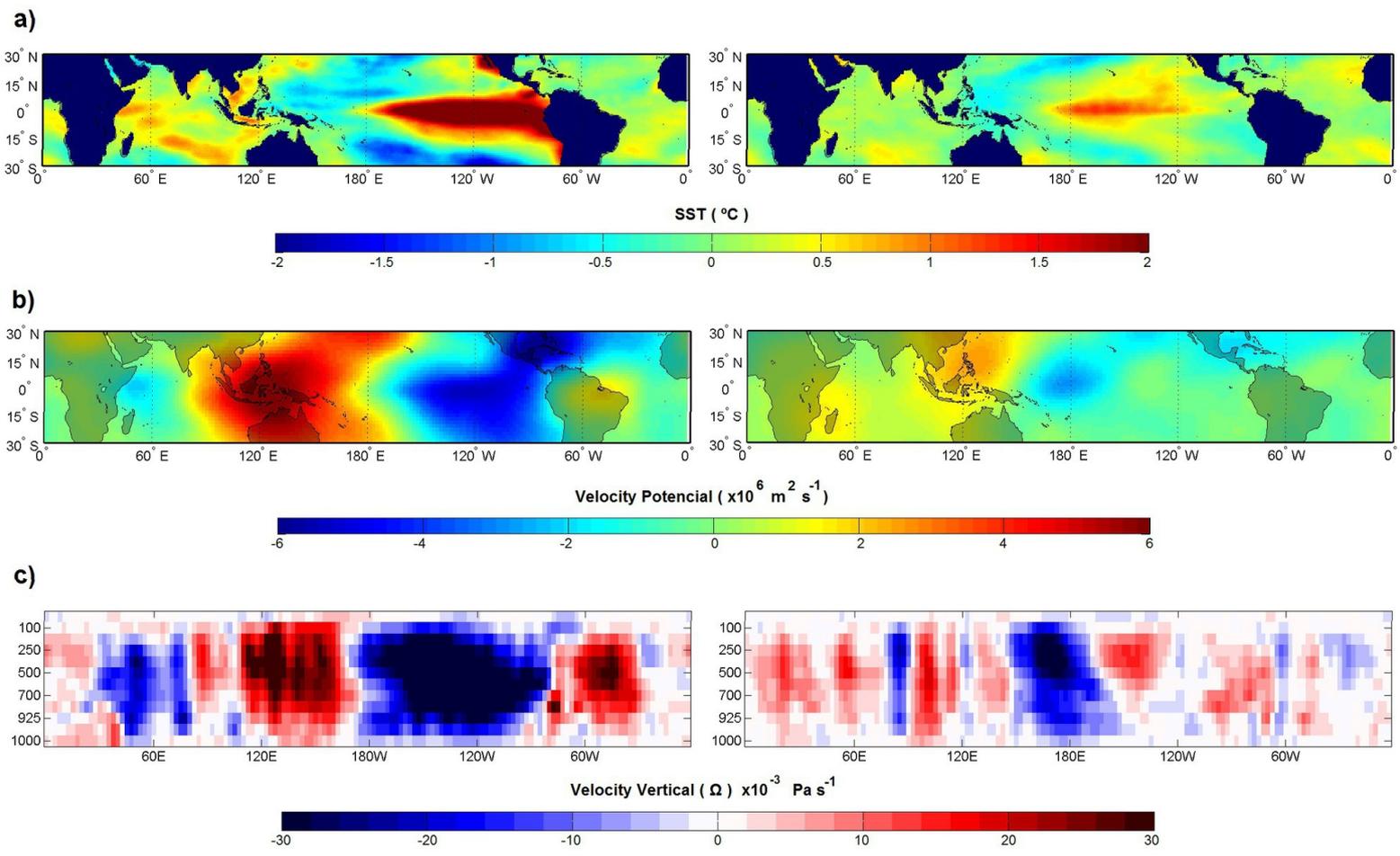

JOC_6047_figure6.jpg

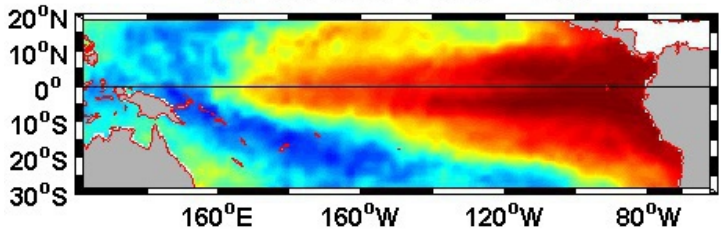
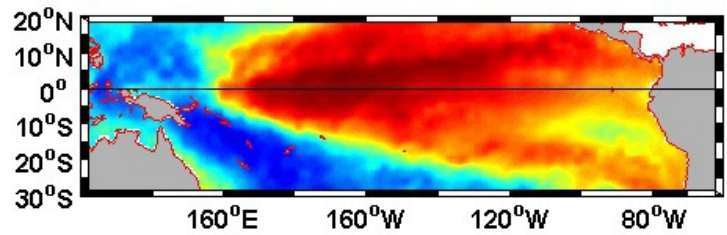
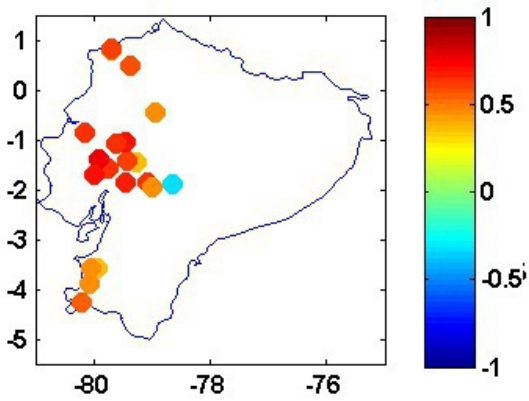
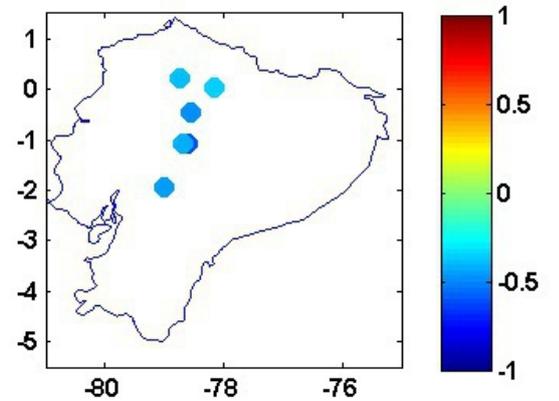
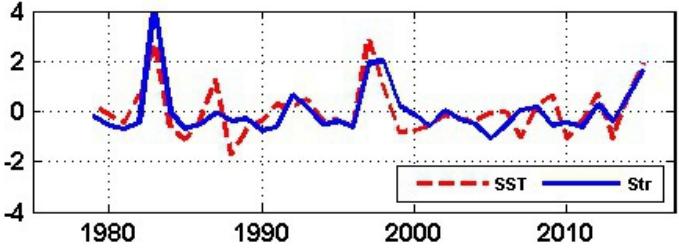
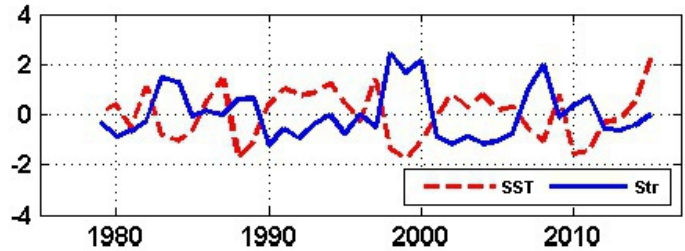

JOC_6047_figure7.jpg

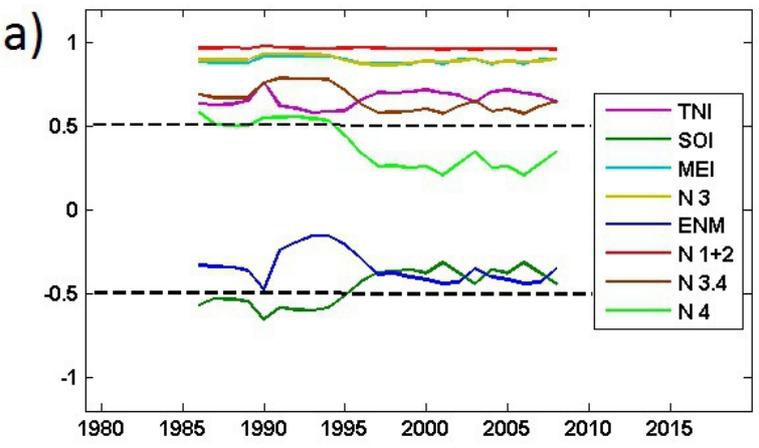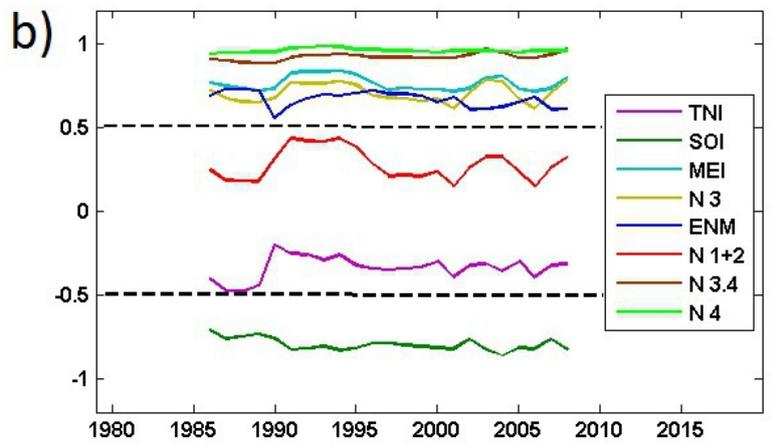

JOC_6047_figure8.jpg

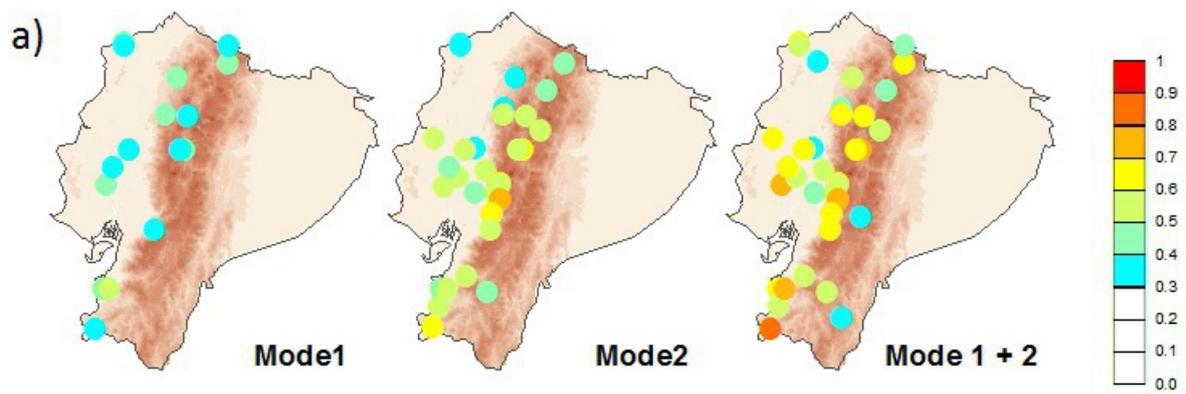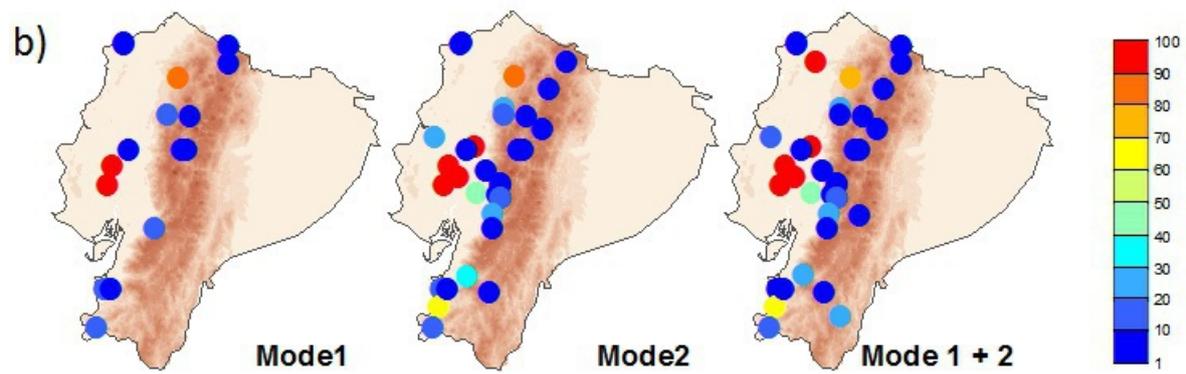

JOC_6047_figure9.jpg

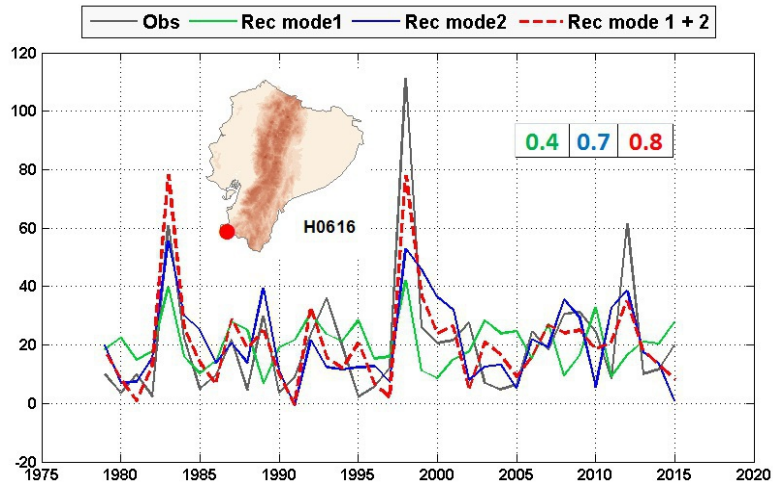
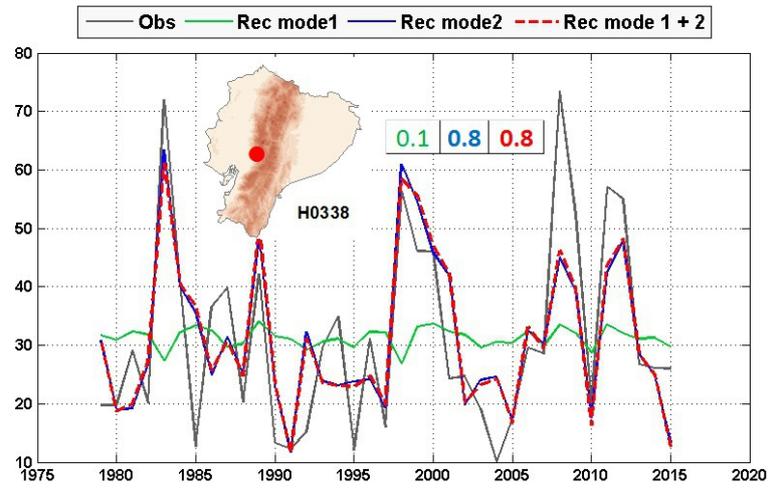
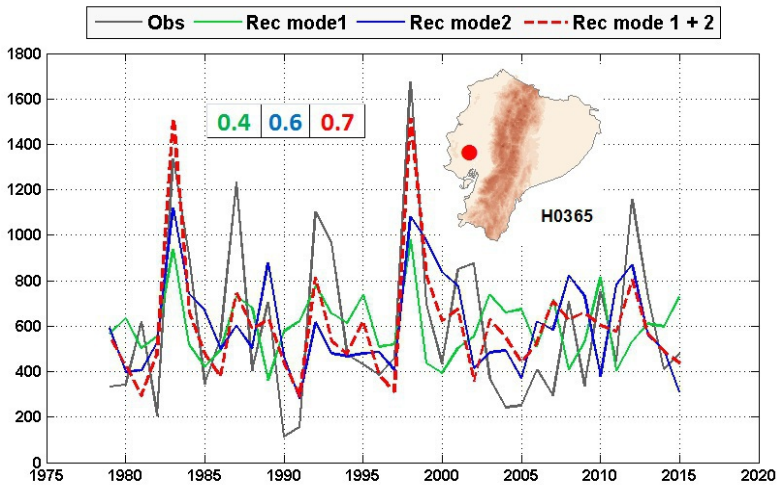
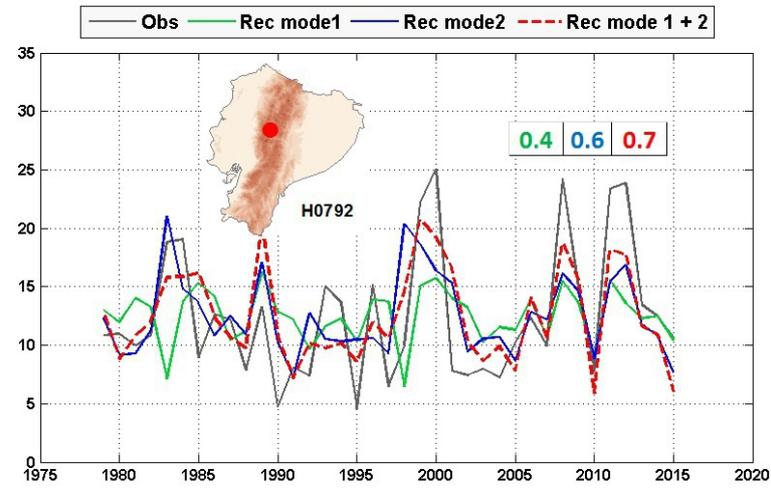
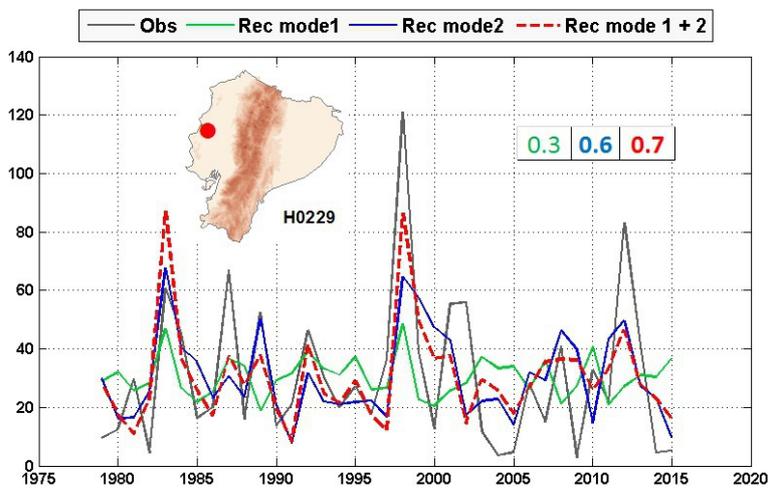
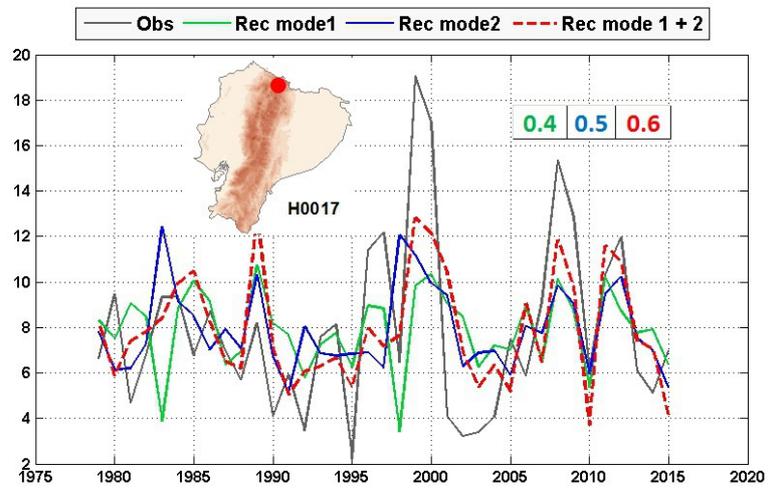

JOC_6047_figure10.jpg

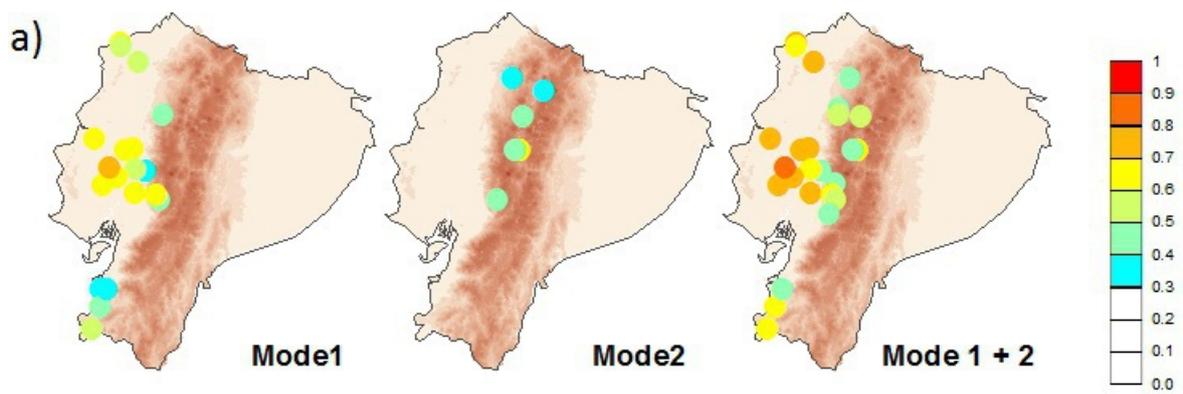
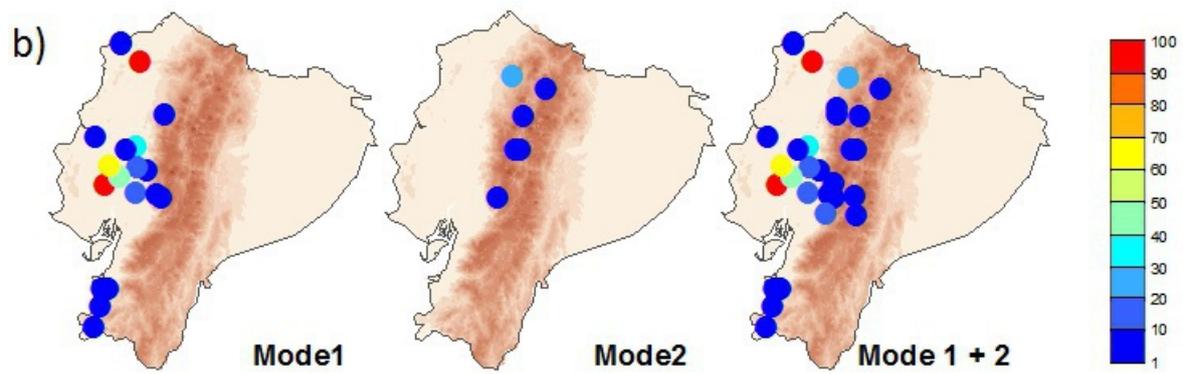

JOC_6047_figure11.jpg

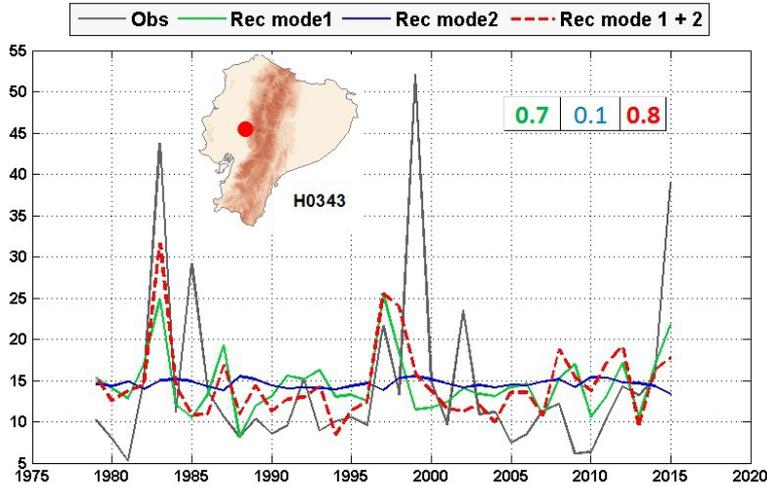
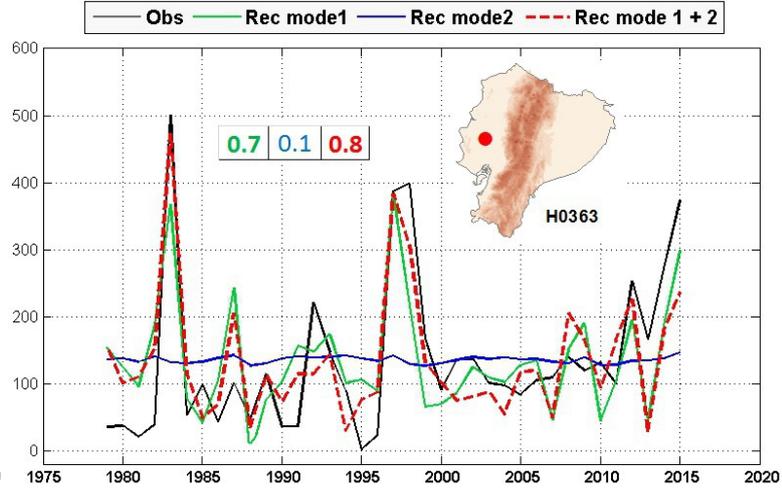
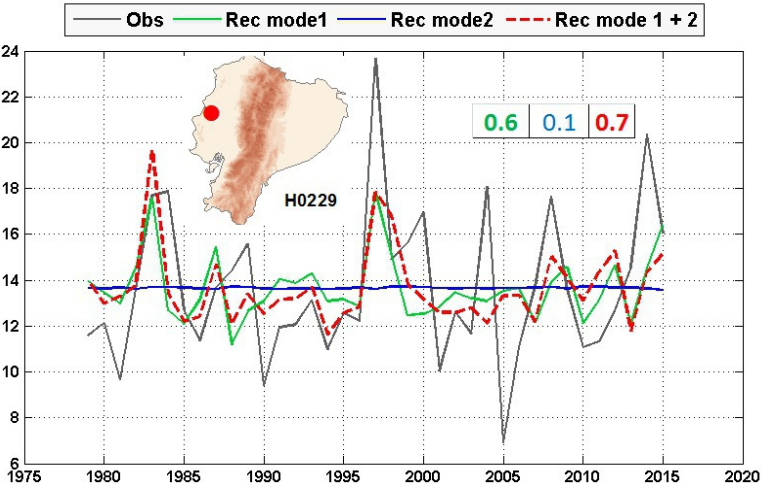
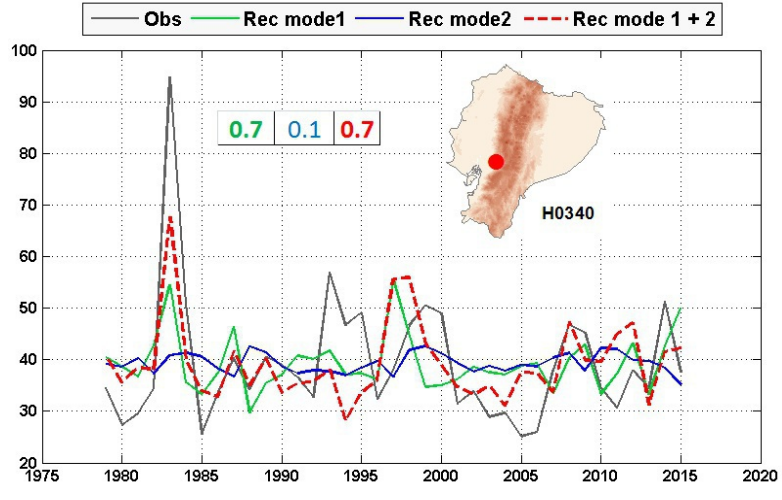
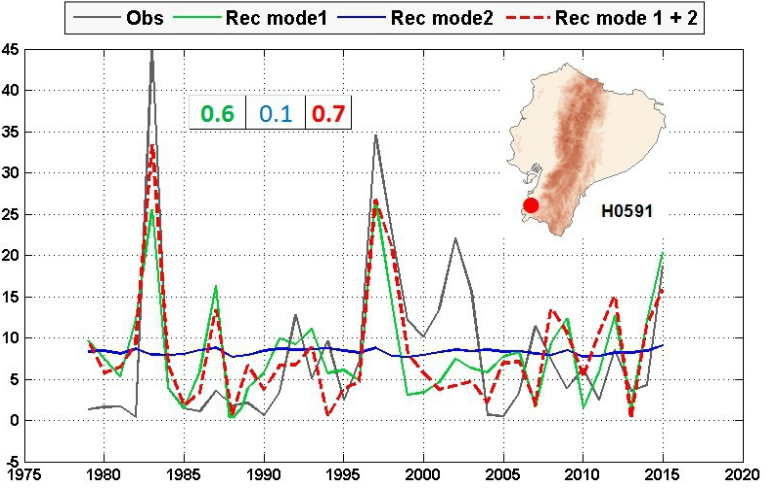
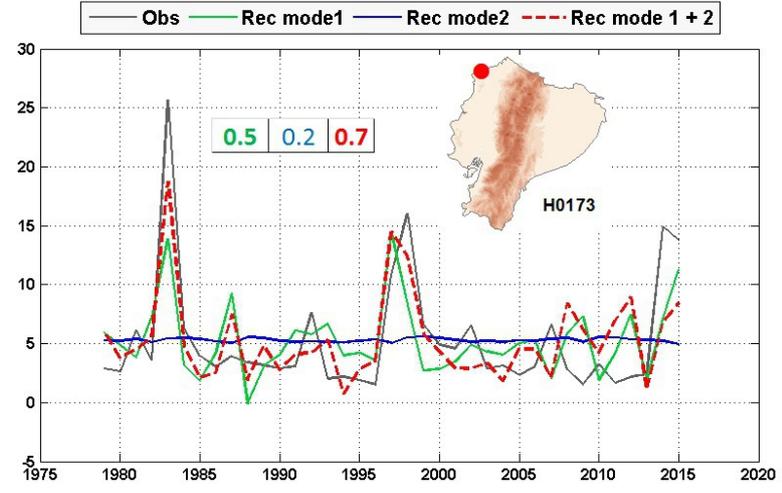

JOC_6047_figure12.jpg

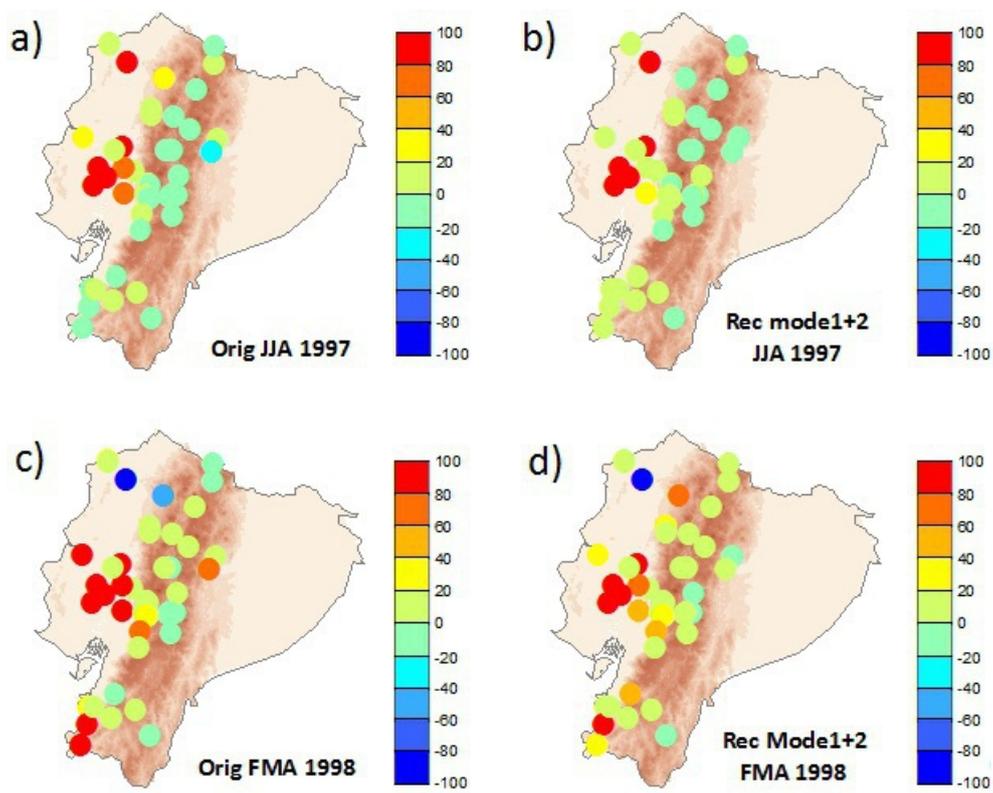

JOC_6047_figure13.jpg

Table 1. Streamflow stations used in this study.

| ID | Basin River | Station Name | Monthly Average (m³/s) | Altitude (masl) | ID | Basin River | Station Name | Monthly Average (m³/s) | Altitude (masl) |
|---|---|---|---|---|---|---|---|---|---|
| H0017 | Mira | Apaqui | 8.92 | 2423 | H0363 | Guayas | Daule | 219.7 | 30 |
| H0091 | Carchi | Grande Jativa | 1.66 | 3.192 | H0365 | Guayas | Daule | 301.4 | 13 |
| H0143 | Esmeraldas | Granobles | 5.55 | 2714 | H0371 | Guayas | Palmar | 63.01 | 15 |
| H0146 | Esmeraldas | Guayllabamba | 126.5 | 679 | H0471 | Cañar | Raura | 12.93 | 984 |
| H0159 | Esmeraldas | Machachi | 3.93 | 2740 | H0530 | Jubones | Jubones | 46.24 | 282 |
| H0161 | Esmeraldas | T. Pilaton | 39.68 | 827 | H0573 | S. Rosa | S. Rosa | 2.75 | 80 |
| H0166 | Esmeraldas | T. Las Pampas | 19.9 | 1110 | H0574 | Arenillas | Arenillas | 8.05 | 20 |
| H0168 | Esmeraldas | Sade | 871.4 | 51 | H0587 | Puyango | Pindo | 29.1 | 90 |
| H0172 | Esmeraldas | Teaone | 5.63 | 16 | H0591 | Puyango | Puyango | 86.71 | 300 |
| H0173 | Esmeraldas | Teaone | 9.18 | 9 | H0616 | Chira | Alamor | 7.81 | 247 |
| H0229 | Chone | Carrizal | 14.25 | 47 | H0625 | Jubones | Alamor | 2.15 | 1080 |
| H0331 | Guayas | Chimbo | 5.7 | 2405 | H0720 | Napo | Misahualli | 17.95 | 800 |
| H0332 | Guayas | Canal | 0.24 | 2419 | H0721 | Napo | Jatunyacu | 275.7 | 570 |
| H0333 | Guayas | San Lorenzo | 1.41 | 2438 | H0722 | Napo | Yanahurco | 2.04 | 3606 |
| H0334 | Guayas | De Chima | 1.58 | 2100 | H0783 | Pastaza | Ozogoche | 2.62 | 3756 |
| H0337 | Guayas | Chimbo | 4.83 | 1480 | H0787 | Pastaza | Alao | 8.01 | 3200 |
| H0338 | Guayas | Pangor | 17.8 | 1452 | H0788 | Pastaza | Puela | 13.73 | 2475 |
| H0340 | Guayas | Bucay | 38.44 | 297 | H0789 | Pastaza | Guargualla | 4.55 | 2828 |
| H0343 | Guayas | Echeandia | 22.75 | 425 | H0790 | Pastaza | Cebadas | 20.03 | 2840 |
| H0346 | Guayas | Zapotal | 146.9 | 40 | H0792 | Pastaza | Cutuchi | 9.95 | 2582 |
| H0347 | Guayas | Quevedo | 201.9 | 68 | H0793 | Pastaza | Nagsiche | 1.63 | 2962 |
| H0348 | Guayas | Vinces | 224.4 | 41 | H0889 | Santiago | Zamora | 79.96 | 902 |
| H0352 | Guayas | Macul | 5.52 | 54 | | | | | |

Table 2. Percentage of stations showing negative, positive and significant (at 95% confidence level) trends in monthly streamflow.

|  | Jan | Feb | Mar | Apr | May | Jun | Jul | Aug | Sep | Oct | Nov | Dec |
|---|---|---|---|---|---|---|---|---|---|---|---|---|
| ( - ) Trend | 22 | 22 | 36 | 33 | 29 | 33 | 44 | 31 | 34 | 33 | 36 | 36 |
| ( + ) Trend | 49 | 64 | 58 | 60 | 64 | 58 | 42 | 33 | 33 | 31 | 40 | 47 |
| ( - ) Sig. 95% | 0 | 0 | 2 | 0 | 2 | 0 | 0 | 13 | 29 | 20 | 13 | 2 |
| ( + ) Sig. 95% | 27 | 11 | 4 | 7 | 4 | 9 | 13 | 22 | 13 | 16 | 11 | 16 |

Table 3. Correlation values between the DJF El Niño indices and the first two DJF expansion coefficients series of tropical Pacific SST, during 1979-2015. Significant correlation values at the 95% confidence level are shown in bold.

|         | DJF SST MODE 1 | DJF SST MODE 2 |
|---------|----------------|----------------|
| TNI     | -0.32          | **-0.92**      |
| SOI     | **-0.87**      | -0.10          |
| MEI     | **0.96**       | 0.08           |
| Niño 3  | **0.97**       | -0.10          |
| ENMI    | **0.53**       | **0.85**       |
| Niño 1+2| **0.83**       | **-0.42**      |
| Niño 3.4| **0.97**       | 0.16           |
| Niño 4  | **0.83**       | **0.57**       |

Table 4. List of years corresponding to the canonical El Niño and El Niño Modoki events used in the composite analysis. The indicated event refers to the winter (DJF) season. Extreme ENSO events have been selected based on the El Niño 3 and El Niño Modoki indices as those years in which winter has an index value roughly equal to or greater than 0.5 standard deviation. For the composites analysis only years without coincidence of the two types of El Niño events were considered (years in black).

| canonical El Niño | El Niño Modoki |
|---|---|
| 1982/1983 | 1979/1980 |
| 1986/1987 | 1986/1987 |
| 1991/1992 | 1990/1991 |
| 1994/1995 | 1991/1992 |
| 1997/1998 | 1992/1993 |
| 2002/2003 | 1994/1995 |
| 2009/2010 | 2002/2003 |
| 2014/2015 | 2004/2005 |
|  | 2009/2010 |
|  | 2014/2015 |

Table 5. As Table 3 for JJA El Niño indices and the first two expansion coefficients series of tropical Pacific JJA SST.

|         | JJA SST MODE 1 | JJA SST MODE 2 |
|---------|----------------|----------------|
| TNI     | **0.62**       | -0.31          |
| SOI     | **-0.54**      | **-0.78**      |
| MEI     | **0.90**       | **0.79**       |
| N 3     | **0.91**       | **0.71**       |
| ENMI    | -0.29          | **0.66**       |
| Niño 1+2| **0.97**       | 0.32           |
| Niño 3+4| **0.69**       | **0.91**       |
| Niño 4  | **0.48**       | **0.94**       |